\documentclass[twocolumn,showpacs,preprintnumbers,amsmath,amssymb]{revtex4-2}

\usepackage{graphicx}
\usepackage{soul}
\usepackage{dcolumn}
\usepackage{xcolor}
\usepackage[colorlinks]{hyperref}
\usepackage[capitalise]{cleveref}
\usepackage[normalem]{ulem}
\DeclareMathOperator{\arccot}{arccot}
\usepackage{bm}

\newcommand{\pd}[2]{ \frac{ \partial #1}{ \partial #2 } }

\newcommand{\Vt}{V_0}
\newcommand{\const}{a}

\newcommand{\edit}[1]{#1}

\begin{document}


\title{Morphological attractors in natural convective dissolution}
\author{Jinzi Mac Huang$^{1,2}$}
\email{machuang@nyu.edu}
\thanks{The authors contributed equally.}
\author{Nicholas J. Moore$^3$}%
 \email{nickmoore83@gmail.com}
\thanks{The authors contributed equally.}
\affiliation{1. NYU-ECNU Institute of Physics and Institute of Mathematical Sciences, New York University Shanghai, Shanghai, 200122, China \\
2. Applied Math Lab, Courant Institute, New York University, New York, NY 10012, USA \\
3. Mathematics Department, United States Naval Academy, Annapolis, MD 21402, USA
}%

\date{\today}

\begin{abstract}
Recent experiments demonstrate how a soluble body placed in a fluid spontaneously forms a dissolution pinnacle — a slender, upward pointing shape that resembles naturally occurring karst pinnacles found in stone forests. This unique shape results from the interplay between interface motion and the natural convective flows driven by the descent of relatively heavy solute. Previous investigations suggest these structures to be associated with shock-formation in the underlying evolution equations, with the regularizing Gibbs-Thomson effect required for finite tip curvature. Here, we find a class of exact solutions that act as attractors for the shape dynamics in two and three dimensions. Intriguingly, the solutions exhibit large but finite tip curvature without any regularization, and they agree remarkably well with experimental measurements. The relationship between the dimensions of the initial shape and the final state of dissolution may offer a principle for estimating the age and environmental conditions of geological structures.
\end{abstract}



\maketitle


Ever-changing geological features on this planet never fail to capture our imagination and inspire new scientific advances. Often, striking features appear when fluid and solid interact, ranging from centimeter scale pebble stones \cite{ristroph2012sculpting, Huang2015} to the kilometer scale karst terrains \cite{sweeting2012karst, ford2013karst}. Even  planetary-scale plate tectonics are believed to have such a fluid-structure interaction origin \cite{Whitehead1972, whitehead1988fluid, Zhong2005, mac2018stochastic}. 

\begin{figure*}[htb]
 \includegraphics[height=1.5in]{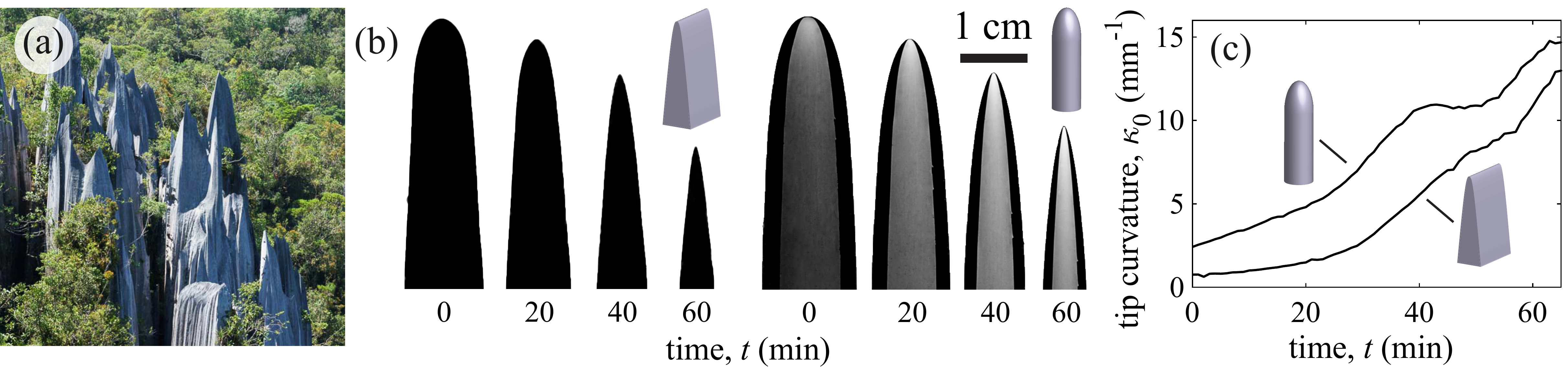}
  \caption{Dissolution-induced sharpening. (a) Limestone structures form the stone forests of Borneo (Grant Dixon). 
  (b)-(c) Dissolution of lab-scale planar and axisymmetric objects unveils the sharpening process; \edit{images from the same set of experiments reported in \cite{mac2020ultra}.
  The observed noise in the curvature measurements results from surface impurities, like bubbles, affecting image tracking. The final radius of curvature at the tip was measured to be 60 $\mu$m.}}
\label{fig1}
\end{figure*}

The direct study of geophysical structures presents unique challenges owing to the vast range of scales, along with the limitation of only seeing the current state. On the other hand, laboratory-scale experiments combined with judicious physical models have proven valuable in explaining certain formations \cite{Meakin2010, dodds2000scaling}, like the growth of icicles \cite{short2006free}, river meandering \cite{Callander1978, Stolum1996, Braudrick2009}, the formation of stalactites and stalagmites \cite{allison1923growth, Short2005, Short2005a}, meteor ablation \cite{amin2019role}, and plate tectonics \cite{Whitehead1972, whitehead1988fluid, Zhong2005, mac2018stochastic}.
In this letter, we investigate one such geomorphological problem, namely the formation of karst pinnacles \cite{gines2009karst,sweeting2012karst}. We will demonstrate the unusual shape dynamics that result in convergence to a morphological attractor.

Commonly seen in South Asia and the island of Madagascar \cite{song1986origination,veress2008origin}, \cref{fig1}(a) shows the typical shape of the karst pinnacles that comprise stone forests. While their origins remain unclear, studies have related such pinnacles to the dissolution process \cite{gines2009karst,ford2013karst,slabe2016laboratory}, as many of these rocks were once immersed under water, and the rock material is slightly water-soluble. Two questions naturally arise: How does the rock evolve into individual pinnacles? Why does each pinnacle exhibit the common feature of a sharp apex?

Aimed at addressing such questions, recent experiments employed lab-scale soluble objects to recreate the stone forests purely from the perspective of dissolution and fluid dynamics \cite{mac2020ultra}. These experiments show stone forests to manifest from a single porous, soluble block, highlighting the sharpening of each karst pinnacle as the key to such formations. In these and other \cite{Nakouzi2014} experiments, no external flow is imposed, rather the transport of relatively heavy solute sustains a natural convective flow that drives shape evolution.

Huang {\it et al.}~2020 and Pegler \& Wykes 2020 proposed a boundary-layer based model capable of predicting sharpening \cite{mac2020ultra, pegler2020shaping}. Notably, the model reduces to a single integro-PDE that governs shape evolution, denoted here as the sharpening equation (SE). Initial numerical evidence and scaling analysis of the SE suggested shock formation and finite-time blowup of the tip curvature \cite{mac2020ultra}. Likewise, similarity solutions of a matched-asymptotic approximation predict unbounded growth of tip curvature for certain initial conditions \cite{pegler2020shaping, pegler2021convective}.

\edit{
\cref{fig1} shows experimental images of dissolving planar and axisymmetric bodies (see \cite{mac2020ultra} for experimental details).
}
Measurements of the tip curvature indeed increase over time, as seen in \cref{fig1}(b) and (c), but interestingly give no clear indication of singular behavior. To reconcile these observations, previous studies appealed to the thermodynamic Gibbs-Thomson (GT) effect \cite{aaron1970effects, perez2005gibbs}, which regularizes the SE and limits the curvature growth.
\edit{However, the strength of the GT term used in previous simulations was at the high end of the range estimated from physical considerations (1-10 $\mu$m) \cite{mac2020ultra}, thus calling into question whether this term accurately modeled a physical effect or was simply acting to regularize the numerics.
For context, the experiments show in \cref{fig1} reach a final tip radius of 60 $\mu$m, suggesting that the GT effect is secondary.
As such, fundamental questions remain: Does the SE support geometric shock formation? Is there a blowup in tip curvature, and, if so, is the blowup only limited in practice by microscale thermodynamics?}

Here, we resolve these and other questions by finding a class of exact solutions to the SE in 2D and 3D that serve as attractors for the shape dynamics. The solutions exhibit large, but finite, tip curvature, indicating that the GT effect is not needed to regularize sharpening. Improved numerical methods, specially tailored to the hyperbolic nature of the SE, show how initially convergent characteristics bend to avoid crossing and eventually straighten in pursuit of the attracting morphology. Revisited experiments confirm the convergence to these exact solutions, thus raising the possibility of using the solutions to infer properties of natural structures.

{\em The model}.---In accordance with Fick's law, a soluble interface retreats with normal velocity proportional to the gradient of the solute field $V_n\propto \nabla c \cdot \mathbf{n}$ \cite{Huang2015, moore2017riemann, mac2021stable}. These dynamics can be greatly complicated by the presence of a fluid flow, which significantly distorts the field $c$ and alters local gradients. The flow may be forced externally \cite{ristroph2012sculpting, rycroft2016, mitchell2017, hewett2017, derr2020, quaife2018boundary, chiu2020viscous, ladd2020dissolution} or driven by buoyancy variations \cite{Nakouzi2014, wykes2018self, mac2020ultra}, as in the present study. The evolution of flow, solute, and body shape are thus inextricably linked.

\begin{figure}[t]
 \includegraphics[width=3.4in]{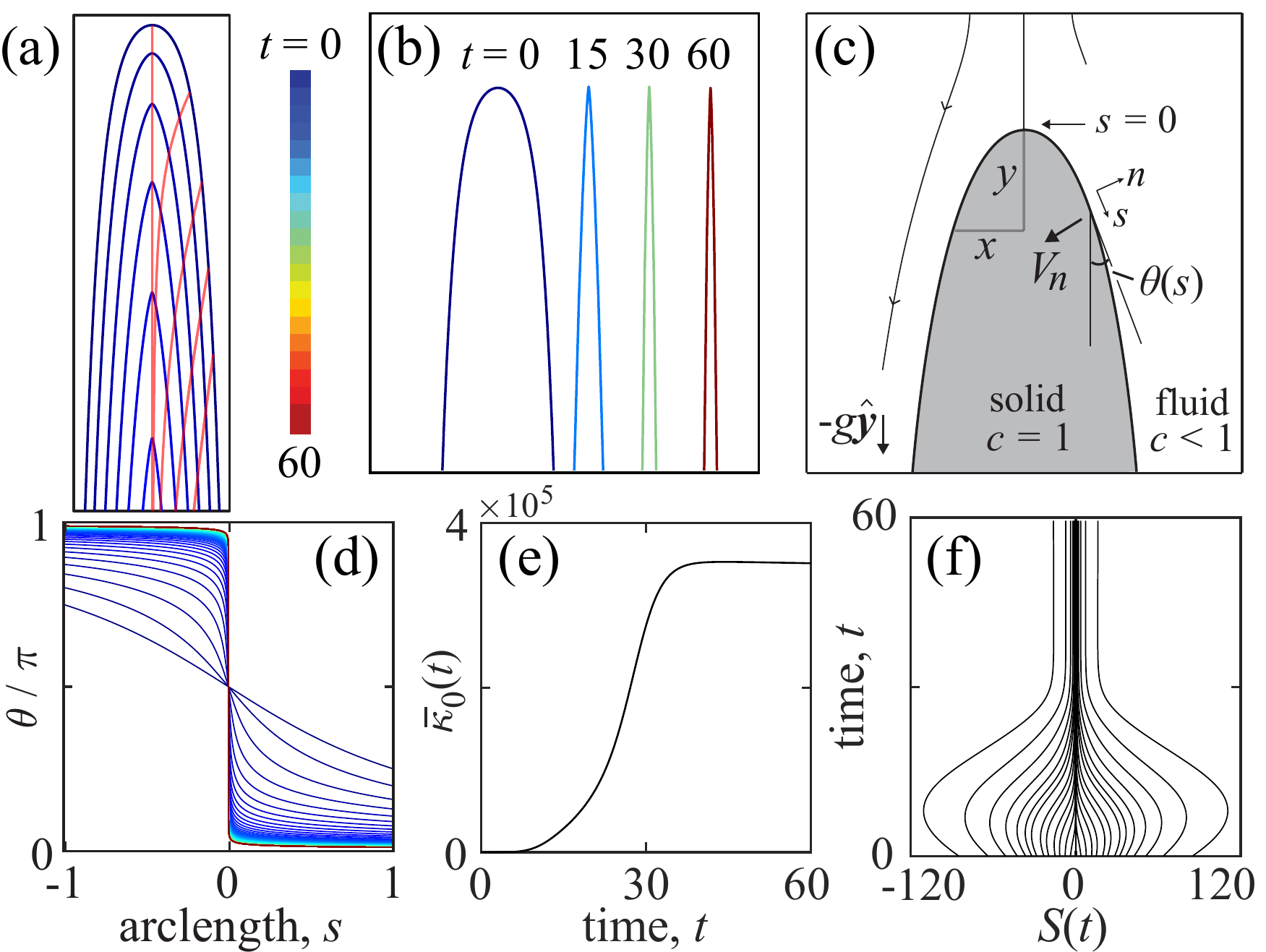}
  \caption{
  Simulating dissolution-induced sharpening. (a) Evolution of the initial shape $\theta = \arccot{s/\ell}$ in 2D. (b) Zooming-in near the apex illustrates the strong sharpening effect. (c) Model schematic. 
  (d) Profiles of the tangent-angle $\theta(s,t)$ show a steep gradient develop near the apex, $s=0$, consistent with (e) a tip curvature that increases by 5 orders of magnitude. (f) Characteristic curves show contours of constant $\theta$, with the physical trajectories shown in (a) with red.
  }
\label{fig2}
\end{figure}

Due to the large Schmidt and Grashof numbers ($\text{Sc} \sim 10^3$ and $\text{Gr} \sim 10^9$, see SI) of the pinnacle experiments, these convective flows are confined to narrow boundary layers, enabling an explicit expression for the 2D interface velocity \cite{schlichting2016boundary, mac2020ultra}:
\begin{equation}
\label{VnEQN}
    V_n = - \const \cos^{\frac{1}{3}}\theta \,
    \left( \int_0^s \cos^{\frac{1}{3}}\theta \, ds' \right)^{-\frac{1}{4}}
\end{equation}
where the surface tangent angle $\theta = \theta(s,t)$ is parameterized by the arclength $s$ from the apex, as illustrated in \cref{fig2}(c). The constant $\const \approx 10^{-7} \mbox{  m}^{5/4}/\mbox{s}$ contains all material and fluid properties. For simplicity, we focus on the 2D case in this Letter, with analogous analysis for axiysymmetric (3D) objects available in the SI.

The $\theta$-$L$ formulation \cite{hou1994removing, alben2002drag, moore2013self, mac2021stable} offers a single, {\em scalar} equation that fully describes shape evolution:
\begin{equation}
\label{thetaEQN}
\pd{\theta}{t} = \pd{V_n}{s} + V_s \pd{\theta}{s}.
\end{equation}
As above, $\theta$ represents the surface tangent angle, and the Cartesian coordinates can easily be recovered from $\frac{d}{ds}(x, y) = (\sin \theta, \cos \theta)$. The artificial tangential velocity $V_s = \int_0^s V_n \partial_s \theta \, ds'$ enforces an invariant metric with respect to arclength, thereby separating $s$ and $t$ as {\em independent} variables. \Cref{thetaEQN} with interface velocity \cref{VnEQN} is the nonlinear integro-PDE proposed in \cite{mac2020ultra}, here called the sharpening equation (SE); see \cite{pegler2020shaping} for the Cartesian counterpart .

Previous investigations employed a finite-difference scheme to solve \cref{thetaEQN}, but with the GT regularization required to maintain numerical stability \cite{mac2020ultra}. Other studies employed a matched-asymptotic expansion, but with approximation error that may grow large with time \cite{pegler2020shaping, pegler2021convective}. In contrast, we introduce a method to directly propagate characteristics of \cref{thetaEQN}, with no regularization and no additional model approximation made. 

To that end, consider a location $s = S^{(0)}$ on the initial geometry, with tangent angle $\Theta^{(0)} = \theta(S^{(0)},0)$. The trajectory $S(t)$ evolves via the ODE:
\begin{equation}
    \label{Sdot}
\dot{S}(t) = \left( R \, \pd{V_n}{s} - V_s\right) \Big\rvert_{s=S(t)}, \quad S(0) = S^{(0)},
\end{equation}
where $R = -(\partial \theta/\partial s)^{-1} = \kappa^{-1}$ is the radius of curvature. Combining \cref{thetaEQN,Sdot} shows that the tangent angle remains constant along such a \textit{characteristic}, $\theta(S(t), t) = \Theta^{(0)}$, thus providing an implicit solution for any initial profile $\Theta^{(0)} = \theta(S^{(0)},0)$. This is the essence of the method of characteristics.

A PDE-based interpretation of \cref{Sdot} is also possible via implicit functions. That is, regard $s = s(\theta,t)$, where $\theta \in (0,\pi)$ is now the independent variable, to obtain:
\begin{align}
\label{sEQN}
& \pd{s}{t} = -\pd{V_n}{\theta} - V_s \, , \\
\label{newVn}
& V_n =  - \const \cos^{\frac{1}{3}} \theta
\left( \int^{\pi/2}_\theta R(\theta', t) \cos^{\frac{1}{3}} \theta' d\theta' \right)^{-\frac{1}{4}} \, ,
\end{align}
where now $R(\theta, t) = -\partial s/\partial \theta$ and $V_s = \int_{\pi/2}^\theta V_n(\theta')d\theta'$. Crucially, the reformulation in terms of $\theta$ implies increased numerical tip-resolution in proportion to the sharpening.
We thus solve \Cref{sEQN,newVn} numerically (see the SI for implementation details) for a class of left-right symmetric initial conditions. 

{\em Results}.---As a first numerical test, we simulate the dissolution of the initial profile $\theta(s, 0) = \arccot(s/\ell)$, with $\ell=1$ and $\const = 1$ (for other values, time could be rescaled by the factor $\ell^{5/4}/a$). As seen in \cref{fig2}(a), dissolution causes the apex to sharpen as the body retreats downwards and diminishes in size. \Cref{fig2}(b) shows a few representative shapes at different stages of dissolution, illustrating the dramatic sharpening effect.
\Cref{fig2}(d) shows the corresponding distributions of the tangent angle, $\theta(s,t)$. Here, a rapid change of tangent angle develops at the tip, as is consistent with the increasing curvature $\kappa = -\partial\theta/ \partial s$ there. Indeed, the rescaled tip curvature $\bar{\kappa}_0(t) = \kappa_0(t)/\kappa_0(0)$ shown in \cref{fig2}(e) increases by 5 orders of magnitude before saturating. 

\Cref{fig2}(f) shows the characteristic curves $(t, S(t))$ corresponding to different constant values of the tangent angle $\theta = \Theta^{(0)}$ [the physical trajectories of these curves are shown in red in \cref{fig2}(a)].
Near the tip ($S\approx0$) characteristics initially converge towards one another, implying a large range of tangent angles crowded into a small region, i.e.~sharpening. Previous discretizations of \cref{thetaEQN} interpreted this convergence as a {\em crossing} of characteristics and thus the formation of a geometric shock. The reformulated \cref{sEQN}, however, reveals that characteristics bend away from one another before ever crossing, thus preventing a finite-time blowup of curvature. Characteristics farther  from the tip are seen to change their direction of travel, initially propagating outwards, and then inwards, before they ultimately straighten and travel vertically. At late times, all characteristics are seen to travel vertically, suggesting that a terminal shape has arrived.

{\em Exact solutions}.---
To examine the possibility of a terminal shape, we take a $\theta$-derivative of \cref{sEQN} to obtain an evolution equation for the radius of curvature \cite{wettlaufer1994geometric, Meakin2010}:
\begin{equation}
\label{equilibrium}
    \pd{R}{t} = V_n + \pd{^2 V_n}{\theta^2}.
\end{equation}
Clearly, a steady-state of \cref{equilibrium} is given by
\begin{equation}
\label{translating}
    V_n = -\Vt{} \sin{\theta}
\end{equation} 
for any constant $\Vt$, which is the recessional rate of the tip. 
\edit{
\cref{translating} represents steady translation of a fixed shape. It is the only steady-state $V_n$ that satisfies left-right symmetry.
}
Inserting \cref{translating} into \cref{newVn} and inverting gives the equilibrium distribution of $R$,
\begin{equation}
\label{R}
    \frac{R^*}{R^*_0} = \frac{1+2\cos^2\theta}{\sin^5\theta} \, .
\end{equation}
This class of equilibrium solutions has one degree of freedom $R^*_0$, which is the equilibrium radius of curvature at the tip. Exact expressions for the Cartesian coordinates of this surface, along with solutions for the corresponding axisymmetric (3D) problem, are given in the SI. Though differences exist in the $\theta$--$L$ formulation of the 2D and 3D problems, the final equilibrium solutions are identical when written in Cartesian coordinates.


\begin{figure}[t]
 \includegraphics[width=2.8in]{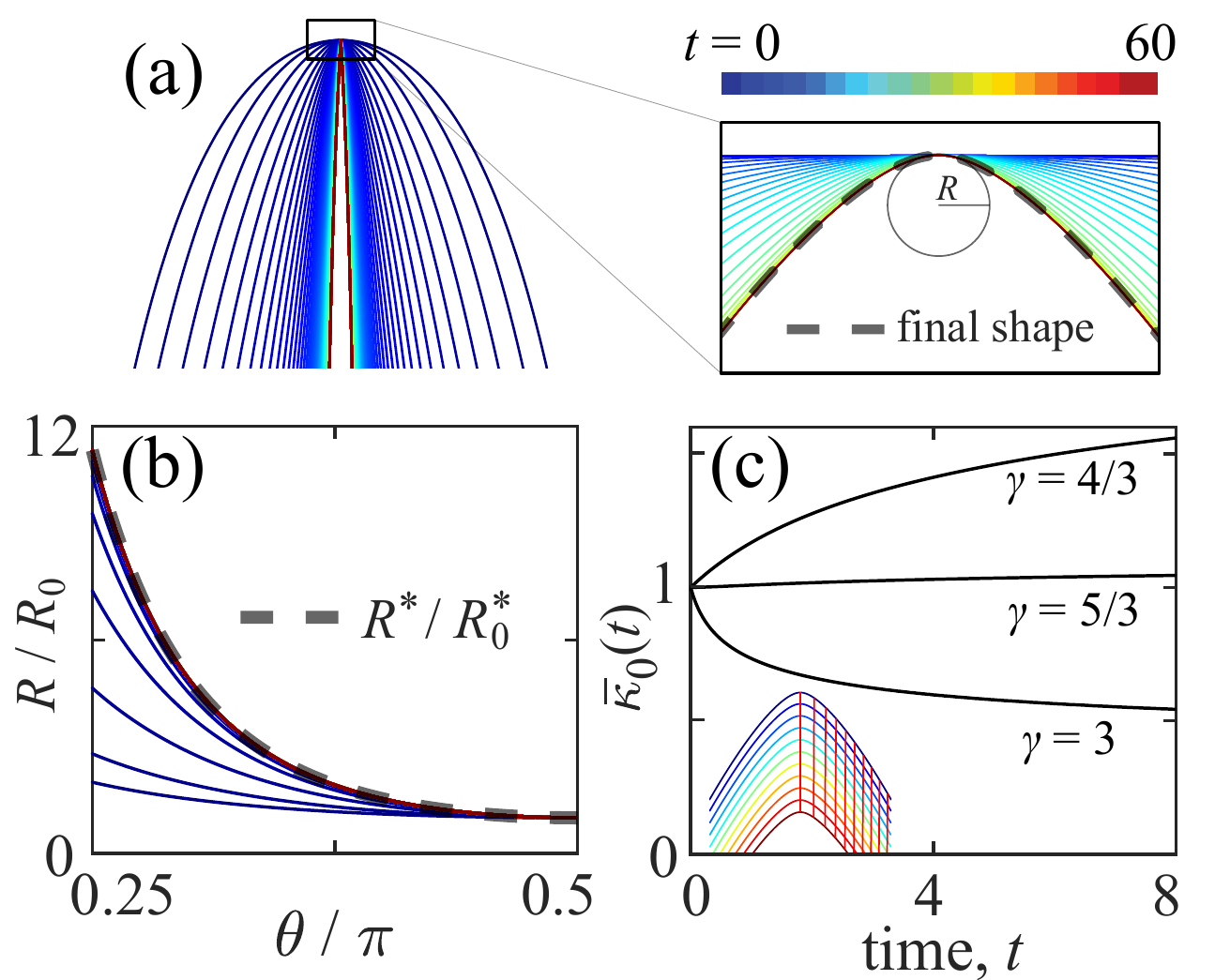}
  \caption{
Convergence towards the equilibrium morphology.
(a) \textit{Left}: Overlaying interfaces in \cref{fig2}(a) shows a common shape to emerge at late times. \textit{Right}:  Zooming-in near the apex further reveals the convergence towards an equilibrium. (b) The rescaled radius of curvature $R/R_0$ tends to the exact distribution predicted by \cref{R}.
(c) Choosing initial shapes near the equilibrium can lead to sharpening or blunting, as predicted by \cref{RvsTime}. \textit{Inset}: physical shape evolution of the case $\gamma=5/3$ reveals straight characteristic paths. 
}
\label{fig3}
\end{figure}

To test the convergence to this final shape, \cref{fig3}(a) shows the simulated interfaces from the previous example, but shifted to have the same apex. As seen here and in the close-up, the interfaces indeed collapse to a single profile at late times. \Cref{fig3}(b) shows that the corresponding distributions of rescaled curvature-radius, $R(\theta,t)/R_0(t)$, converge to the equilibrium shape \cref{R}.

Having observed the convergence to the predicted morphology, several questions remain: What happens for different initial conditions?
What determines the final tip radius
$R^*_0 = \lim_{t\to\infty}R(0,t)$? And how can the results be reconciled with previous  infinite-curvature predictions \cite{mac2020ultra, pegler2021convective}? 
To address these questions, we consider a local expansion in small $w = \cos \theta$:
\begin{equation}
s(w,t) = a_1(t) w + a_3(t) w^3 + \dots
\end{equation}
where odd-symmetry has been used. Thanks to the change of variables, $V_n$ can be calculated exactly for any power $w^n$. Inserting into \cref{sEQN} produces, at leading order, $\dot{R}_0 \propto - R^{-1/4}_0$, which is consistent with \cite{mac2020ultra} and predicts finite-time blowup of curvature. However, retaining the higher-order terms gives
\begin{equation}
\label{RvsTime}
    \dot{R}_0 \propto - R^{-1/4}_0 \left(1-\frac{3}{5}\gamma \right), \quad \gamma(t) = \frac{a_3(t)}{a_1(t)} \, ,
\end{equation}
which is an exact relation (no truncation). \Cref{RvsTime} opens the possibility for the curvature divergence to be controlled by the term $\left(1-\frac{3}{5}\gamma \right)$, and indeed the equilibrium solution \cref{R} has the property $\gamma = 5/3$.

To further examine this possibility, \cref{fig3}(c) shows the simulated dissolution of three initial conditions surrounding the equilibrium: $s(w,0) = a_1(0) w + a_3(0) w^3$, with $\gamma$ initially set to $5/3$, $4/3$,  and $3$. The figure confirms that $\gamma = 5/3$ results in nearly constant curvature \footnote{The curvature does not remain exactly constant due to the fact that this initial condition is not the terminal shape from \cref{R}, but a truncation of it.}, whereas $\gamma > 5/3$ (resp.~$\gamma < 5/3$) leads to decreasing (resp.~increasing) curvature, consistent with the sign of $\dot{R}_0$ in \cref{RvsTime}. Thus, both tip sharpening and blunting are possible \cite{pegler2020shaping, pegler2021convective}, with the value of $\gamma$ determining which occurs. 
The case $\gamma=4/3$ leads to tip sharpening, but, due to the proximity to the equilibrium, not nearly as much as in our first numerical example. Thus, the enormous curvature growth observed in \cref{fig2} should not always be expected, as it depends on the initial shape.

\begin{figure}[h]
 \includegraphics[width=3in]{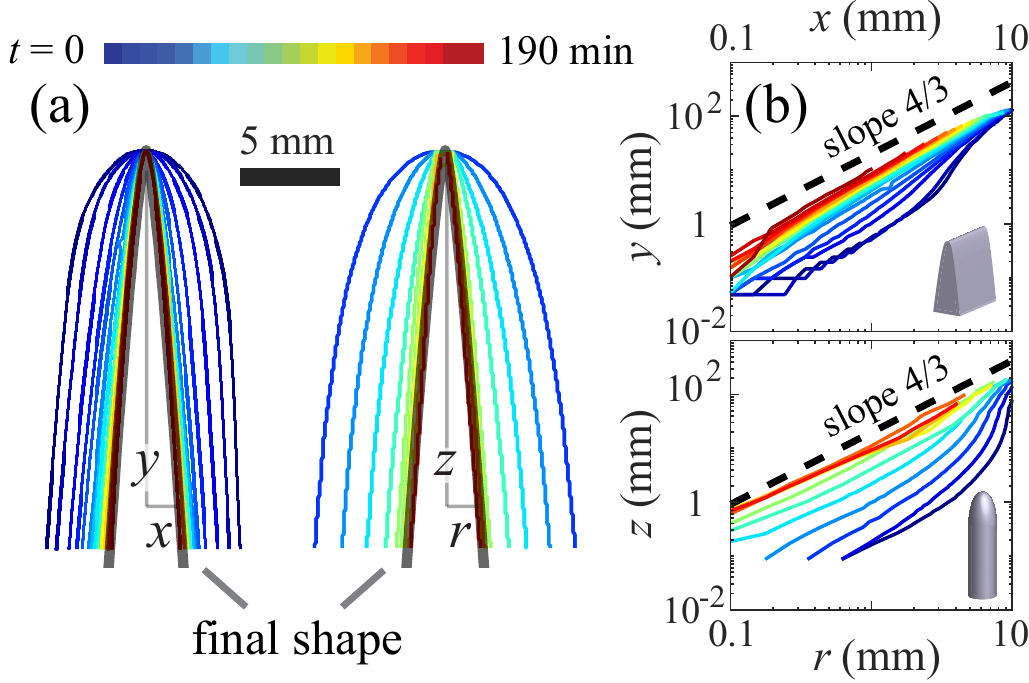}
  \caption{
Comparison with laboratory experiments \edit{shown in Fig.~1}. (a) When overlaid, the profiles measured from the planar (2D) and axisymmetric (3D) experiments are seen to converge to the shape predicted by \cref{R}. Plotting the experimentally measured surface coordinates on a log-scale confirms the far-field prediction \cref{finalShape-2}.
  }
\label{fig4}
\end{figure}

We now turn attention to the experimentally-measured shapes that were shown in \cref{fig1}, for planar (2D) and axisymmetric (3D) geometries. \Cref{fig4}(a) compares the experimental profiles (shifted to have the same apex) to the equilibrium morphology of \cref{R} (thick gray curve). At late times, the experimental profiles all collapse onto the predicted shape in both 2D and 3D, thus conclusively confirming that \cref{R} accurately describes the equilibrium spire-morphology of a body dissolving under its own solute-induced convective flow. This agreement with laboratory experiments also validates modeling assumptions made, including the boundary-layer and quasi-steady approximations and the omission of GT effects.

A second test is made possible by the far-field ($|s| \to \infty$) behavior of the equilibrium solution
\begin{align}
\label{finalShape-2}
    y &\sim \frac{3}{4} \, {R_0^*}^{-1/3} \, x^{4/3} \, ,
\end{align}
which holds in both 2D and 3D [with $(x,y) \to (r,z)$ in 3D, see SI].
\Cref{fig4}(b) shows a log-scale comparison between this predicted 4/3-power law and the experimental measurements. At late times, the experimental profiles indeed converge to the predicted power law in both cases.
We note that this power-law is consistent with one of the similarity solutions found in \cite{pegler2020shaping, pegler2021convective}, which would need to be asymptotically matched to an inner (near-tip) solution. Those similarity solutions, however, predict continued evolution of shape, whereas we have found convergence to a final form. \edit{Numerical and experimental evidence suggests this final morphology to be a stable attractor. }

\edit{Closer examination of our exact solutions offers an interpretation of the flow-physics underlying the convergence in shape dynamics. Within the boundary layer, a few competing effects exist. First, the apex is in contact with nearly pure liquid, whereas a solute mixture washes over the downstream portions. In isolation, this effect would cause the apex to retreat fastest. On the other hand, the buoyancy-driven flow accelerates as it advances downstream, due to the accumulation of dense solute as well as the increase in surface steepness. This effect enhances convection-induced dissolution on {\em downstream} portions. Which effect is stronger depends on the detailed geometry of the object, and it is the interplay between the two that drives shape change. Ultimately, balance is achieved by the steadily-translating distribution, $V_n = -\Vt{} \sin{\theta}$, which shows that the dissolution-rate is highest at the tip ($\theta=\pi/2$) and decreases locally in proportion to the surface steepness. At this stage, the mass loss rate of the pinnacle has a simple scaling $dm/dt \sim 2\int_{\theta_0}^{\pi/2} V_n(\theta) R^*(\theta) \, d\theta \sim -V_0 R_0^*\sim - (R_0^*)^{3/4}$, implying that the mass loss slows as the tip sharpens.
}


{\em Discussion.}---
In this Letter, we have described, in exact form, the final spire morphology of a body being reshaped under its own dissolution-induced natural convective flow, thereby concluding the search from \cite{Nakouzi2014, wykes2018self, mac2020ultra, pegler2020shaping, pegler2021convective}. 
Carefully designed numerics show that, rather than forming a geometric shock, characteristics avoid crossing to pursue this terminal shape, which exhibits large, but finite tip curvature.
\edit{This situation is perhaps analogous to exact solutions found in the context of free surface flows, whose finite curvature reversed previous hypotheses on the formation of cusp singularities \cite{Moffatt1992}.} 


The simple, explicit nature of our solutions suggests that they may be used to infer properties, e.g. age or past environmental conditions, of natural structures. To take one example, suppose that a karst pinnacle at time $t_0$ has height $h(t_0)$, width $d(t_0)$, and that its apex dissolves at the rate $\dot{h}(t_0) = V(t_0)$. As it nears the final shape,  \cref{finalShape-2} suggests $h(t)d(t)^{-4/3} = h(t_0)d(t_0)^{-4/3}$ and the constant tip velocity gives $h(t) - h(t_0) = V(t_0)(t-t_0)$. These two relationships comprise a closed system for $(d(t), h(t))$ at any given time -- including the past ($t<t_0$) and the future ($t>t_0$)-- thus offering the potential to estimate the past dimensions or, if the dimensions can be estimated through other means, the age of the structure. To take this idea one step further, the typical spacing $L$ between pinnacles in a stone forest approximates the initial width $d(0) \approx L$, thus offering simple estimates for the pinnacle's initial height $h(0) = h(t_0)[L/d(t_0)]^{4/3}$ and its age $t_0 = [h(t_0) - h(0)] / V(t_0)$.
Though natural systems involve a range of other complicating factors \edit{(such as rainfall, turbulent boundary layers, and fracture)} our calculations, based principally on dissolution and fluid dynamics, may offer a leading-order understanding of these amazing structures.

\edit{Aspects of our analysis can be extended to other physical systems. For example,  \cref{equilibrium} can have a separable solution $R(\theta,t) = A(\theta)B(t)$, corresponding to the self-similar evolution of erodible and soluble bodies immersed in an externally forced flow \cite{ristroph2012sculpting, moore2013self, Huang2015}.
Meanwhile, our approach can be applied to dynamics with an opposite sign in $V_n$, seen in growing systems like crystallization \cite{wettlaufer1994geometric} and the formation of stalactites \cite{Short2005, Short2005a}. }

\bibliography{manuscript}

\end{document}


\preprint{APS/123-QED}

\title{Supplementary Information: Morphological attractors in natural convective dissolution}
\author{Jinzi Mac Huang$^{1,2}$}
\author{Nicholas J. Moore$^3$}%
\affiliation{1. NYU-ECNU Institute of Physics and Institute of Mathematical Sciences, New York University Shanghai, Shanghai, 200122, China \\
2. Applied Math Lab, Courant Institute, New York University, New York, NY 10012, USA \\
3. Mathematics Department, United States Naval Academy, Annapolis, MD 21402, USA
}%

\date{\today}
\maketitle

\section{Governing Equations and Boundary layer approximation}

\begin{figure}[h]
 \includegraphics[width=3.5in]{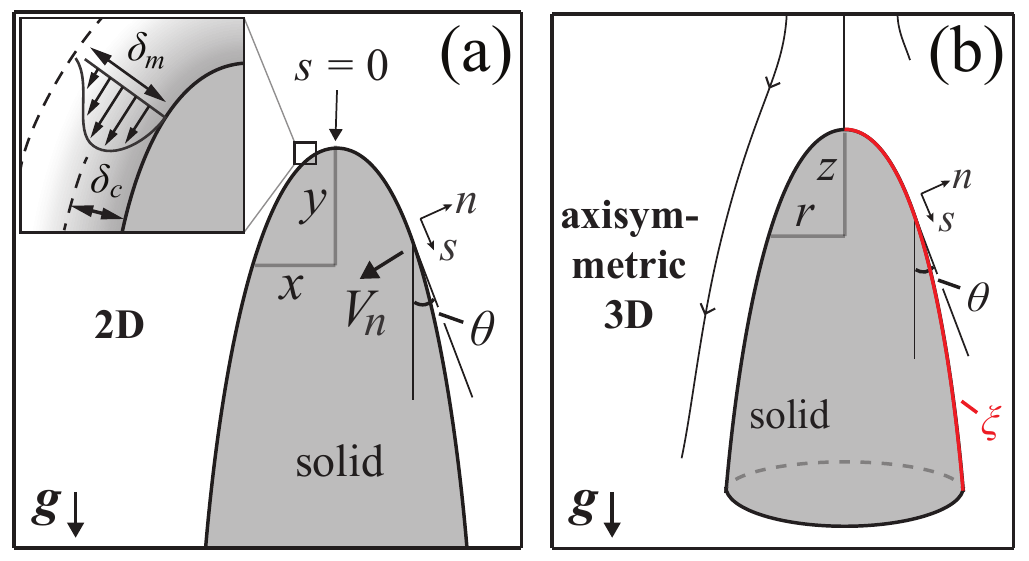}
  \caption{Schematic of the dissolution model. (a) Dissolution of 2D objects: The boundary $\Gamma$ is symmetric with respect to $x=0$, and a boundary layer structure (inset) develops near the solid-liquid interface. (b) Dissolution in axisymmetric 3D: The boundary $\Gamma$ is defined by revolving an arc $\xi$ around the axis $r=0$, and a similar boundary layer structure exists. }
\label{fig1-supp}
\end{figure}

The dissolution configurations for both the 2D and axisymmetric 3D objects are shown in \cref{fig1-supp}. In the fluid domain, there are two equations governing the dissolution process: the advection-diffusion equation for the concentration field $c$, and the Navier Stokes equation for the fluid velocity field $\mathbf{u}$ and pressure field $p$. Using the Boussinesq approximation of the density-stratified Navier-Stokes equations \cite{tritton2012physical, camassa2012}, we obtain 
\begin{align}
\label{fluid}
&\frac{\partial c}{\partial t} + \mathbf{u}\cdot\nabla c = D \Delta c,\\
&\frac{\partial\mathbf{u}}{\partial t} + \mathbf{u}\cdot\nabla \mathbf{u} = - \frac{1}{\rho_l} \nabla p + \nu \Delta \mathbf{u}  + \beta c \mathbf{g},\\
&\nabla\cdot \mathbf{u} = 0 ,
\label{flow}
\end{align}
with boundary conditions, 
\begin{equation}
\label{bc}
c|_\Gamma=c_{s}, \qquad \mathbf{u}|_\Gamma=\mathbf{0}, \qquad \textrm{and} \qquad
V_n = (\beta+1)^{-1} D \mathbf{n}\cdot \nabla c|_\Gamma.
\end{equation}

For the dissolution of crystalized sugar, the physical constants are fluid kinematic viscosity $\nu \sim 10^{-6}$ m$^2/$s, solute diffusivity $D \sim 10^{-9}$ m$^2/$s, acceleration due to gravity $g = |\mathbf{g}| = 9.8$ m$/$s$^2$, and the density contrast between the solid ($\rho_s$) and the liquid ($\rho_l$),  $\beta = (\rho_s-\rho_l)/\rho_l\sim 1$. At the boundary, the concentration reaches its saturated value of $c_s=1$ and the velocity satisfies a no-slip condition $\mathbf{u}|_\Gamma=\mathbf{0}$, assuming that interface motion is slow compared to the fluid flow. The interface velocity $V_n$ in the outward normal direction $\mathbf{n}$ is set by the concentration gradient $\nabla c$, due to the conservation of mass and Fick's law of diffusion \cite{moore2017riemann}.

Dimensionlessly, $\ScN = \nu / D $ defines the ratio of fluid viscosity to molecular diffusivity, and the Grashof number $\GrN = g \beta L^3 /\nu^2$ defines the ratio of buoyancy to viscous forces \cite{tritton2012physical,schlichting2016boundary} given a typical body size $L$. The physical parameters of sugar dissolution, $\ScN\sim 10^3$ and $\GrN\sim 10^9$, lead to the development of boundary layer structures \cite{schlichting2016boundary}, as shown in the inset of \cref{fig1-supp}(a). Boundary layers of momentum and concentration fields have their typical sizes of $\delta_m$ and $\delta_c$. The distribution of flow velocity $\mathbf{u}$ and solute concentration $c$ within this boundary layer is well known \cite{mac2020ultra, schlichting2016boundary}, enabling explicit expressions for $\nabla c$ and the interface velocity $V_n$ as parameterized by arc length $s$ from the tip:
\begin{gather}
\label{vn1}
    V_n(s,t) = - a \frac{\cos^{1/3}\theta(s,t)}{\left[ \int_0^s \cos^{1/3}\theta(s',t) ds' \right]^{1/4}}     \quad \mbox{in 2D,}\\
    V_n(s,t) = - a \frac{r(s,t)^{1/3} \cos^{1/3}\theta(s,t)}{\left[\int_0^{s} r(s',t)^{4/3} \cos^{1/3}\theta(s',t)ds' \right]^{1/4}} \quad \mbox{in axisymmetric 3D.}
\end{gather}

The constant $a = 0.503(\beta+1)^{-1} (\frac{g \beta D^3}{\nu} )^{1/4} \approx  10^{-7} \ \mbox{m}^{5/4}/\mbox{s}$ contains only material parameters, and $r(s,t) = \int_0^s \sin{\theta(s',t)} ds'$ is the radius of revolution as shown in \cref{fig1-supp}(b). Here we have applied Mangler and Saville-Churchill transformations; for details, see pp. 265-273 and pp. 323-324 of Schlichting's \textit{Boundary-Layer Theory} \cite{schlichting2016boundary}.

The $\theta$-$L$ framework \cite{hou1994removing, moore2013self, quaife2018boundary, chiu2020viscous, mac2021stable} provides an the evolution equation for the tangent angle $\theta(s,t)$,
\begin{equation}
\label{thetaEQN}
\pd{\theta}{t} = \pd{V_n}{s} + V_s \pd{\theta}{s}.
\end{equation}
Here the artificial tangent velocity $V_s = \int_0^s V_n \partial_s \theta \, ds'$ is added to separate $s$ and $t$ as independent variables.

As shown in the text, it is convenient to regard the tangent angle $\theta$ as a dependent variable, and evolve the distribution of arclength $s(\theta,t)$ to obtain the shape dynamics. In this case, $\theta = \pi/2$ is the apex of dissolving object so $s(\pi/2,t) = 0$. Defining $R(\theta,t) = - \partial s/\partial \theta$ as the radius of curvature and rewriting the revolution radius as $r(\theta,t) = \int^{\pi/2}_{\theta} R(\theta',t) \sin{\theta'}  d\theta'$, we have
\begin{gather}
    \label{vn2}
    V_n(\theta,t) = - a \frac{\cos^{1/3}\theta}{\left[ \int^{\pi/2}_{\theta} R(\theta',t)  \cos^{1/3}\theta' d\theta' \right]^{1/4}}     \quad \mbox{in 2D,}\\
    \label{vn3}
    V_n(\theta,t) = - a \frac{r(\theta,t)^{1/3} \cos^{1/3}\theta}{\left[\int^{\pi/2}_{\theta} r(\theta',t)^{4/3} R(\theta',t)\cos^{1/3}\theta' d\theta' \right]^{1/4}} \quad \mbox{in axisymmetric 3D.}
\end{gather}
The evolution equation in this reformulated framework is 
\begin{align}
\label{dsdt}
& \pd{s}{t} = -\pd{V_n}{\theta} - V_s \, ,
\end{align}
with $V_s = \int_{\pi/2}^\theta V_n(\theta')d\theta'$.

\section{Numerical Methods}


We now introduce the numerical method to simulate  convection-driven dissolution in 2D. We seek to solve \cref{dsdt} with normal velocity given by \cref{vn1} or equivallently \cref{vn2}. We introduce the function 
\begin{equation}
G = \cos^{-4/3} \theta \, \int_0^{s} \cos^{1/3} \theta(s') ds'
\end{equation}
so that the normal velocity can be written as $V_n = -\const G^{-1/4}$. The function $G$ can be rewritten as 
\begin{equation}
\label{Geq}
G = \cos^{-4/3} \theta \int_0^s s'^{1/3} g(s') ds'
\end{equation}
where
\begin{equation}
\label{geq}
g(s) = s^{-1/3} \cos^{1/3}\theta(s) 	\, .
\end{equation}
The tip curvature is given by $\kappa_0 = -\theta_s(0)$, and  taking the limit $s \to 0$ of \cref{geq} gives the special value $g(0) =  \kappa_0^{1/3}$. Thus, $g(s)$ is perfectly well behaved near the tip, and so the term $s'^{1/3}$ in \cref{Geq} represents the true non-smoothness of the integrand. The integrand of \cref{Geq}, while not singular itself, has a singular first derivative, which would limit the accuracy of a standard quadrature method.

We can remove this first-derivative singularity as follows:
\begin{align}
G &= \cos^{-4/3} \theta(s) 
\left( \int_0^s s'^{1/3} \left( g(s')-g(0) \right) ds' + g(0) \int_0^s s'^{1/3} ds' \right)  \\
&= \cos^{-4/3} \theta(s) \left( \int_0^s s'^{1/3} \left( g(s')-g(0) \right) ds' + \frac{3}{4} g(0) s^{4/3} \right),
\end{align}
thus giving a new integrand with a continuous first derivative (and singular second derivative) for which a standard quadrature method will suffice. We perform quadrature on the function
\begin{align}
\label{Heq}
H = G - \frac{3}{4} g(0) s^{4/3} \cos^{-4/3}(\theta)  
= \cos^{-4/3} \theta \int_0^s s'^{1/3} \left( g(s') - g(0) \right) ds'.
\end{align}

We discretize uniformly in the tangent angle, $\theta_n = \pi/2 - n \Delta \theta$, producing an arclength discretization $s_n = s(\theta_n)$ that possesses increased resolution in regions of high curvature, e.g.~the apex. Applying the trapezoid rule to \cref{Heq} gives the recursive formula for $H_n = H(s_n)$:
\begin{align}
\label{Hrec}
& H_n = r_n^{4/3} H_{n-1} + \left( \frac{s_n - s_{n-1}}{2 w_n} \right)
\left( 1 + r_n^{1/3} - \left( \frac{\kappa_0}{w_n} \right)^{1/3} 
\left(s_{n-1}^{1/3} + s_n^{1/3} \right) \right) \qquad
\end{align}
where $w_n := \cos(\theta_n)$ and $r_n := w_{n-1}/w_{n}$.

The starting value of $H_0 = 0$ is required to apply the recursion. Further, while $H_1$ could in principle be obtained from \cref{Hrec}, a more accurate estimate of $H_1$ can improve the overall accuracy of the scheme (since $H_1$ is nearest to the point of non-smoothness and all subsequent values will depend on $H_1$).
Performing a local expansion in the variable $w = \cos(\theta) \ll 1$ produces:
\begin{align}
H_1 = \frac{3}{20} \left( \frac{s_2 - 2 s_1}{ \Delta \theta} \right) + O(\Delta \theta^4).
\end{align}
Then the remaining values of $H$ can be computed through \cref{Hrec}, and $G$ is calculated via 
\begin{align}
G &= H + \frac{3}{4} g(0) \, s^{4/3} \cos^{-4/3}(\theta). 
\end{align}
The normal and tangential velocities are then calculated through
\begin{align}
& V_n = -a G^{-1/4} \\
\label{Vs}
& V_s = \int_{\pi/2}^\theta V_n(\theta') \, d\theta'.
\end{align}
where the integral in \cref{Vs} is computed with the trapezoid rule on the (uniform) $\theta$-grid. The computations of $V_n$ and $V_s$ are second-order accurate in $\Delta \theta$. 
To solve \cref{dsdt}, we then calculate $\partial{V_n}/\partial{\theta}$ via centered finite differences. This differencing applied to the numerically calculated $V_n$ (which is not infinitely smooth) reduces the accuracy of the computation of the right-hand-side of \cref{dsdt} to 3/2-order in $\Delta \theta$.


Once the right-hand-side of \cref{dsdt} is determined, time stepping is performed with MATLAB's ode15s, which is a variable-step and variable-order method for stiff equations. We choose a stiff solver due to the fact that \cref{dsdt} behaves parabolically near the apex:
\begin{equation}
    \pd{s}{t} = \alpha \pd{^2s}{\theta^2} + O(\cos{\theta}),
\end{equation}
where $\alpha = 3a/\left[28G(0)^{5/4}\right] > 0$. Implicit time stepping would be an alternate method to ensure numerical stability.
In all numerical experiments, we evolve $N=200$ equally-spaced characteristics. We set $a = 1$ as one can always rescale time for different values of $a$. 


\section{Final shapes}

In this section, we further elaborate on the exact solutions that describe a class of equilibrium morphologies. To this end, we differentiate \cref{dsdt} with respect to $\theta$ to obtain an evolution equation for the radius of curvature $R(\theta, t) = -\partial s / \partial \theta$:
\begin{equation}
\label{dRdt}
    \pd{R}{t} = V_n + \pd{^2 V_n}{\theta^2}.
\end{equation}
We seek a steady-state solution to \cref{dRdt}, which can be obtained by setting $V_n = V_0 \sin{\theta}$, where $V_0$ is any constant, physically representing the recessional rate of the tip. These are the only equilibrium distributions of $V_n$ that satisfy left-right symmetry.




\subsection{2D final shape}


In order to evaluate the 2D final shape, we first invert the relationship between $V_n$ and $R$ in \cref{vn2} to obtain 
\begin{equation}
    R(\theta,t) = R_0(t) \left[\sin{(\theta)} V_n(\theta,t) +3\cos{(\theta)}\pd{V_n(\theta,t)}{\theta}\right].
\end{equation}
Inserting  $V_n = V_0 \sin{\theta}$, we obtain the steady-state radius of curvature distribution in 2D 
\begin{equation}
\label{Rsup}
    \frac{R^*}{R^*_0} = \frac{1+2\cos^2\theta}{\sin^5\theta} \, .
\end{equation}

The arc length $s$ and Cartesian coordinates $(x,y)$ satisfy the following differential equations,
\begin{align}
    \frac{ds}{d\theta} &= -R(\theta),\\
    \frac{dx}{d\theta} &= -R(\theta)\sin{\theta}, \\
    \frac{dy}{d\theta} &= -R(\theta)\cos{\theta}.
\end{align}

Inserting the equilibrium distribution \cref{Rsup} and integrating with the boundary conditions $s(\pi/2) = x(\pi/2) =y(\pi/2)=0$ yields the explicit formulas
\begin{align}
    \label{seq}
    \frac{s^*}{R^*_0} &= \frac{\cos{\theta}}{8\sin^2\theta}+\frac{3\cos{\theta}}{4\sin^4\theta}-\frac{1}{8}\ln\left(\tan{\frac{\theta}{2}}\right),\\
    \label{x}
    \frac{x^*}{R^*_0} &= \frac{\cos{\theta}}{\sin^3\theta}, \\
    \label{y}
    \frac{y^*}{R^*_0} &= -\frac{1}{\sin^2\theta}+\frac{3}{4\sin^4\theta}+\frac{1}{4},
\end{align}
which define a one-parameter ($R_0^*$) family of 2D equilibrium shapes. 


There are two limits in which useful, approximate relationships can be obtained through asymptotic expansions of \cref{seq,x,y}.
First, a near-tip ($\theta = \pi/2$) expansion in the variable $w = \cos \theta \ll 1$ gives
\begin{equation}
    \frac{s^*}{R_0^*} = w+\frac{5}{3}w^3 + \frac{12}{5} w^5+\frac{22}{7}w^7+\cdots,
\end{equation}
and thus the equilibrium shape satisfies $\gamma = a_3/a_1 = 5/3$ as discussed in the main text. 
Second, far away from the tip, $\theta\to0$, \cref{x,y} give the asymptotic formulas,
\begin{equation}
    \frac{x}{R^*_0} \sim \theta^{-3}, \quad \frac{y}{R^*_0} \sim \frac{3}{4}\theta^{-4}.
\end{equation}
which imply the far-field relationship
\begin{equation}
y \sim \frac{3}{4} {R_0^*}^{-1/3} x^{4/3}.
\end{equation}
This $4/3$-power law was confirmed by experimental measurements, and, furthermore, may allow one to infer properties of natural structures, as discussed in the main text.


\subsection{Axisymmetric 3D final shape}

In the axisymmetric (3D) setting, similar analysis of \cref{vn3} produces the equilibrium distribution,
\begin{equation}
\label{R3D}
    \frac{R^*}{R^*_0} = \frac{1+2\cos^2\theta}{2\sin^5\theta} -  \frac{\cos{\theta}}{2\sin^4\theta}\frac{d\ln r(\theta)}{d\theta}.
\end{equation}
The cylindrical coordinates $(r,z)$, shown in \cref{fig1-supp}(b), satisfy
\begin{align}
    \frac{dr}{d\theta} &= -R(\theta)\sin{\theta}, \\
    \frac{dz}{d\theta} &= -R(\theta)\cos{\theta}.
\end{align}
Inserting the equilibrium distribution \cref{R3D} and integrating with boundary conditions $r(\pi/2) = z(\pi/2) = 0$ yields the axisymmetric 3D final shape, 
\begin{align}
\label{r}
    \frac{r^*}{R^*_0} &= \frac{\cos{\theta}}{\sin^3\theta}, \\
    \label{z}
    \frac{z^*}{R^*_0} &= -\frac{1}{\sin^2\theta}+\frac{3}{4\sin^4\theta}+\frac{1}{4}.
\end{align}
Comparing \cref{r,z} with \cref{x,y}, we see that 2D and axisymmetric 3D dissolution actually share the same final shape. In the far field ($\theta\to0$), we once again have the relationship
\begin{equation}
z \sim \frac{3}{4} {R_0^*}^{-1/3} r^{4/3}.
\end{equation}

\bibliography{manuscript}